\documentclass[prc,twocolumn,superscriptaddress,showpacs]{revtex4-1}

\usepackage{graphicx}
\usepackage{dcolumn}
\usepackage{bm}
\usepackage{subfigure} 
\begin{document}


\title{First Measurement of the $^{96}$Ru($p, \gamma$)$^{97}$Rh Cross Section for the $p$-Process with a Storage Ring}


\author{Bo Mei}
\affiliation{GSI-Helmholtzzentrum f\"{u}r Schwerionenforschung GmbH, Darmstadt, Germany}
\affiliation{ Goethe University Frankfurt, Germany }

\author{Thomas Aumann}
\affiliation{ Technische Universit\"{a}t Darmstadt, Germany }

\author{Shawn Bishop}
\affiliation{ Technische Universit\"{a}t M\"{u}nchen, Germany }

\author{Klaus Blaum}
\affiliation{ Max-Planck-Institut f\"{u}r Kernphysik, Heidelberg, Germany }

\author{Konstanze Boretzky}
\affiliation{GSI-Helmholtzzentrum f\"{u}r Schwerionenforschung GmbH, Darmstadt, Germany}

\author{Fritz Bosch}
\affiliation{GSI-Helmholtzzentrum f\"{u}r Schwerionenforschung GmbH, Darmstadt, Germany}

\author{Carsten Brandau}
\affiliation{GSI-Helmholtzzentrum f\"{u}r Schwerionenforschung GmbH, Darmstadt, Germany}

\author{Harald Br\"{a}uning}
\affiliation{GSI-Helmholtzzentrum f\"{u}r Schwerionenforschung GmbH, Darmstadt, Germany}

\author{Thomas Davinson}
\affiliation{ University of Edinburgh, United Kingdom }

\author{Iris Dillmann}
\affiliation{GSI-Helmholtzzentrum f\"{u}r Schwerionenforschung GmbH, Darmstadt, Germany}

\author{Christina Dimopoulou}
\affiliation{GSI-Helmholtzzentrum f\"{u}r Schwerionenforschung GmbH, Darmstadt, Germany}

\author{Olga Ershova}
\affiliation{ Goethe University Frankfurt, Germany }

\author{Zsolt F\"{u}l\"{o}p}
\affiliation{ ATOMKI, Debrecen, Hungary }

\author{Hans Geissel}
\affiliation{GSI-Helmholtzzentrum f\"{u}r Schwerionenforschung GmbH, Darmstadt, Germany}

\author{Jan Glorius}
\affiliation{ Goethe University Frankfurt, Germany }

\author{Gy\"{o}rgy Gy\"{u}rky}
\affiliation{ ATOMKI, Debrecen, Hungary }

\author{Michael Heil}
\affiliation{GSI-Helmholtzzentrum f\"{u}r Schwerionenforschung GmbH, Darmstadt, Germany}

\author{Franz K\"{a}ppeler}
\affiliation{ Karlsruher Institut f\"{u}r Technologie, Germany }

\author{Aleksandra Kelic-Heil}
\affiliation{GSI-Helmholtzzentrum f\"{u}r Schwerionenforschung GmbH, Darmstadt, Germany}

\author{Christophor Kozhuharov}
\affiliation{GSI-Helmholtzzentrum f\"{u}r Schwerionenforschung GmbH, Darmstadt, Germany}

\author{Christoph Langer}
\affiliation{ National Superconducting Cyclotron Laboratory, Michigan State University, East Lansing, Michigan, USA }

\author{Tudi Le Bleis}
\affiliation{ Technische Universit\"{a}t M\"{u}nchen, Germany }

\author{Yuri Litvinov}
\affiliation{GSI-Helmholtzzentrum f\"{u}r Schwerionenforschung GmbH, Darmstadt, Germany}

\author{Gavin Lotay}
\affiliation{ University of Edinburgh, United Kingdom }

\author{Justyna Marganiec}
\affiliation{GSI-Helmholtzzentrum f\"{u}r Schwerionenforschung GmbH, Darmstadt, Germany}

\author{Gottfried M\"{u}nzenberg}
\affiliation{GSI-Helmholtzzentrum f\"{u}r Schwerionenforschung GmbH, Darmstadt, Germany}

\author{Fritz Nolden}
\affiliation{GSI-Helmholtzzentrum f\"{u}r Schwerionenforschung GmbH, Darmstadt, Germany}

\author{Nikolaos Petridis}
\affiliation{GSI-Helmholtzzentrum f\"{u}r Schwerionenforschung GmbH, Darmstadt, Germany}

\author{Ralf Plag}
\affiliation{ Goethe University Frankfurt, Germany }
\affiliation{GSI-Helmholtzzentrum f\"{u}r Schwerionenforschung GmbH, Darmstadt, Germany}

\author{Ulrich Popp}
\affiliation{GSI-Helmholtzzentrum f\"{u}r Schwerionenforschung GmbH, Darmstadt, Germany}

\author{Ganna Rastrepina}
\affiliation{ Goethe University Frankfurt, Germany }

\author{Ren{\'e} Reifarth}
\email[]{reifarth@physik.uni-frankfurt.de}
\affiliation{ Goethe University Frankfurt, Germany }

\author{Bj\"{o}rn Riese}
\affiliation{GSI-Helmholtzzentrum f\"{u}r Schwerionenforschung GmbH, Darmstadt, Germany}

\author{Catherine Rigollet}
\affiliation{ KVI-CART, University of Groningen, The Netherlands }

\author{Christoph Scheidenberger}
\affiliation{GSI-Helmholtzzentrum f\"{u}r Schwerionenforschung GmbH, Darmstadt, Germany}

\author{Haik Simon}
\affiliation{GSI-Helmholtzzentrum f\"{u}r Schwerionenforschung GmbH, Darmstadt, Germany}

\author{Kerstin Sonnabend}
\affiliation{ Goethe University Frankfurt, Germany }

\author{Markus Steck}
\affiliation{GSI-Helmholtzzentrum f\"{u}r Schwerionenforschung GmbH, Darmstadt, Germany}

\author{Thomas St\"{o}hlker}
\affiliation{GSI-Helmholtzzentrum f\"{u}r Schwerionenforschung GmbH, Darmstadt, Germany}
\affiliation{Helmholtz-Institut Jena, Jena, Germany}

\author{Tam\'{a}s Sz\"{u}cs}
\affiliation{ ATOMKI, Debrecen, Hungary }

\author{Klaus S\"{u}mmerer}
\affiliation{GSI-Helmholtzzentrum f\"{u}r Schwerionenforschung GmbH, Darmstadt, Germany}

\author{G\"{u}nter Weber}
\affiliation{GSI-Helmholtzzentrum f\"{u}r Schwerionenforschung GmbH, Darmstadt, Germany}
\affiliation{Helmholtz-Institut Jena, Jena, Germany}

\author{Helmut Weick}
\affiliation{GSI-Helmholtzzentrum f\"{u}r Schwerionenforschung GmbH, Darmstadt, Germany}

\author{Danyal Winters}
\affiliation{GSI-Helmholtzzentrum f\"{u}r Schwerionenforschung GmbH, Darmstadt, Germany}

\author{Natalya Winters}
\affiliation{GSI-Helmholtzzentrum f\"{u}r Schwerionenforschung GmbH, Darmstadt, Germany}

\author{Philip Woods}
\affiliation{ University of Edinburgh, United Kingdom }

\author{Qiping Zhong}
\affiliation{GSI-Helmholtzzentrum f\"{u}r Schwerionenforschung GmbH, Darmstadt, Germany}

\begin{abstract}
This work presents a direct measurement of the $^{96}$Ru($p, \gamma$)$^{97}$Rh cross section via a novel technique using a storage ring, which opens opportunities for reaction measurements on unstable nuclei. A proof-of-principle experiment was performed at the storage ring ESR at GSI in Darmstadt, where circulating $^{96}$Ru ions interacted repeatedly with a hydrogen target. The $^{96}$Ru($p, \gamma$)$^{97}$Rh cross section between 9 and 11~MeV has been determined using two independent normalization methods. As key ingredients in Hauser-Feshbach calculations, the $\gamma$-ray strength function as well as the level density model can be pinned down with the measured ($p, \gamma$) cross section. Furthermore, the proton optical potential can be optimized after the uncertainties from the $\gamma$-ray strength function and the level density have been removed. As a result, a constrained $^{96}$Ru($p, \gamma$)$^{97}$Rh reaction rate over a wide temperature range is recommended for $p$-process network calculations.
\end{abstract}


\maketitle

\section{INTRODUCTION}
There are 35 $p$-nuclei on the neutron deficient side of the valley of stability between $^{74}$Se and $^{196}$Hg, which are shielded against production by the neutron-capture processes and are produced in the $p$-process. The main production of these $p$-nuclei occurs via ($\gamma, n$), ($\gamma, p$) and ($\gamma, \alpha$) reactions, and subsequent beta decays in the so-called $\gamma$-process. Therefore, it is essential to determine cross sections of photodisintegration reactions or their inverse reactions for $p$-process network calculations \cite{Arnould2003,Rauscher2013}. Thousands of nuclear reactions are involved in network calculations for $p$-process nucleosynthesis. However, only very few of the required cross sections have been measured by experiments, and thus most of them rely solely on predictions of the Hauser-Feshbach (HF) model codes, e.g., NON-SMOKER \cite{Rauscher2000} and TALYS \cite{Koning2005}, which often have very large uncertainties from nuclear input parameters \cite{Rauscher2013}.

Following current $p$-process network calculations \cite{Rayet1995}, the pattern of the solar abundance for about 60\% of the $p$-nuclei is reproduced within a factor of 3. However, for the $^{92,94}$Mo and $^{96,98}$Ru isotopes, an underproduction of a factor of 20-50 has been calculated for the $\gamma$-process in Core Collapse Supernova models with massive stars of 13-25 M$_{\odot}$ \cite{Arnould2003}. This deficiency has motivated the search for additional production mechanisms, e.g., the $rp$-process \cite{Schatz1998} or the $\nu$$p$-process \cite{Frohlich2006}, but also intensified efforts to remove the uncertainty in required nuclear physics parameters by measuring reaction cross sections. These cross sections are most sensitive to nuclear parameters, i.e., the $\gamma$-ray strength function, nuclear level density and optical potential, in the HF model \cite{Rauscher2013}. However, these critical parameters are not well constrained by experiments, and hence there are large differences between predictions using different parameters.

Most of the existing experimental data for the $p$-process were measured in direct kinematics using stable isotope targets \cite{Rauscher2013}. However, a direct measurement on unstable nuclei is still a major challenge \cite{Rauscher2013}. In this work, we present a novel method using a heavy-ion storage ring developed to measure cross sections of low-energy nuclear reactions, e.g., ($p, \gamma$) reactions, in inverse kinematics for nuclear astrophysics. This method offers some key advantages over traditional methods. It can efficiently use the beam and is well suited for measurements on unstable nuclei with half-lives longer than several minutes. This method was suggested already several decades ago \cite{Rolfs1988,Bertulani1997}. However, it was not realized until the latest achievements in producing, cooling, decelerating, and storing of heavy ions, as well as developments in the nuclear detection system at the experimental storage ring (ESR) of GSI \cite{Steck2004}. This novel technique provides a unique condition for the direct measurement of ($p, \gamma$) reactions around the energy range of astrophysical interest \cite{Kappeler2011} and has been successfully demonstrated for the first time by measuring the $^{96}$Ru($p, \gamma$)$^{97}$Rh cross section between 9~MeV/u and 11~MeV/u. A preliminary data analysis at 11~MeV/u has been reported in conference proceedings \cite{Zhong2010,Rastrepina2011}. Here we report a full analysis with all corrections made. Experimental results of the present work allow us to constrain the most important parameters in the HF model, and thus provide a reliable prediction for this reaction over a wide energy range.

\section{EXPERIMENTAL DETAILS}
During this experiment, $^{96}$Ru ions from the linear accelerator (UNILAC) were first accelerated to 100~MeV/u in the heavy-ion synchrotron (SIS) and then stripped to the bare charge state of 44+ using a 11~mg/cm$^{2}$ carbon stripper foil. The fully stripped ions were injected into the ESR and slowed down to 9~MeV/u, 10~MeV/u  and 11~MeV/u, respectively, by ramping the magnetic fields and the frequency of the radio-frequency (rf) system synchronously. However, large beam losses occurred during this deceleration phase mainly due to imperfections of the ramping parameters. The $^{96}$Ru$^{44+}$ ions were cooled by the electron cooler before and after the slowing down phase to a small diameter (about 5~mm) and momentum spread (around 10$^{-3}$). After the final slowing down phase, about 5$\times$$10^{6}$ $^{96}$Ru$^{44+}$ ions were stored in the ESR with a lifetime of several hundred seconds when the hydrogen target was switched off. Finally, a windowless hydrogen microdroplet target of high density \cite{Kuhnel2009} was switched on and the decelerated $^{96}$Ru$^{44+}$ ions were focused onto this target for nuclear reactions. The great advantage of this storage ring method is that unreacted $^{96}$Ru$^{44+}$ ions were recycled and repeatedly impinged on the hydrogen target for reactions. Considering the revolution frequency of about 400~kHz for $^{96}$Ru$^{44+}$ ions circulating in the ring and the thickness of the H$_{2}$ target of about $10^{13}$~particles/cm$^{2}$, a luminosity of about 2$\times$$10^{25}$~/cm$^{2}$/s has been achieved.

At about 10~MeV/u, the main reaction channels of $^{96}$Ru$^{44+}$ with the hydrogen target include the atomic electron capture (EC) reactions and different nuclear reactions. The former contain mainly two parts, namely the non-radiative electron capture (NRC) and the radiative electron capture (REC) accompanied by the emission of a photon. There are large uncertainties of about 30\% according to Ref.~\cite{Eichler2007} in accurately determining the absolute beam intensities, target densities and the beam-target overlap at the ESR. To remove these large uncertainties, both the K-shell REC (K-REC) products (photons) and the EC products were registered by our detectors, which allowed us to absolutely determine the ($p, \gamma$) cross section by using two normalization methods.

\begin{figure}
\centering
\includegraphics[width=8.6cm]{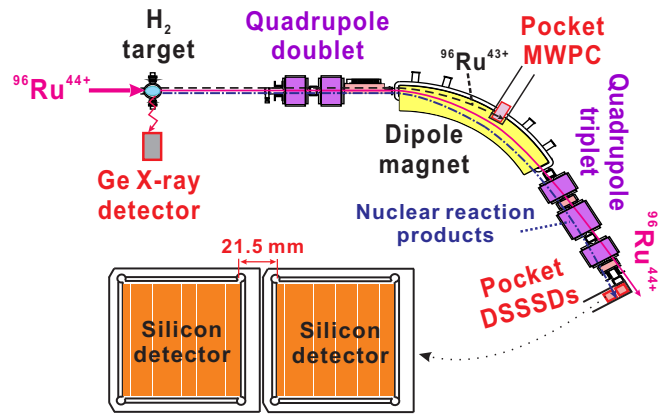}
\caption{(Color online) The experimental setup for proton capture reactions in inverse kinematics at the ESR. The H$_{2}$ target and the detectors used in this experiment are marked. Both the Multi-Wire Proportional Chamber (MWPC) and the double-sided silicon strip detectors (DSSSDs) were placed in pockets with 25~$\mu$m thick windows. A schematic view of the DSSSDs is also shown. The unreacted $^{96}$Ru$^{44+}$ ions (magenta solid line after the target) were recycled and focused on the H$_{2}$ target repeatedly during the measurement phase.}
\label{fig:exp_setup_ESR}
\end{figure}

Fig.~\ref{fig:exp_setup_ESR} shows the experimental setup from the hydrogen target to all used detectors at the ESR.
The X-rays emitted from the atomic reactions were registered by a Ge detector mounted close to the target interaction area at the observation angle of 90$^{\circ}$ with respect to the beam axis. The detection efficiency of this Ge detector was calibrated with mixed $\gamma$-ray sources in the energy range between 10~keV and 140~keV. In the K-REC energy region, an intrinsic efficiency of about 88\% has been reached for this detector with a solid angle ($\Delta \Omega$) of about 2.5$\times$$10^{-3}$~sr. The down-charged $^{96}$Ru ions ($^{96}$Ru$^{43+}$, $^{96}$Ru$^{42+}$, $^{96}$Ru$^{41+}$, etc.) produced by EC were recorded by a position sensitive Multi-Wire Proportional Chamber (MWPC) \cite{Klepper2003} mounted in the vacuum chamber of the first dipole magnet behind the target, as indicated in Fig.~\ref{fig:exp_setup_ESR}. This detector was operated with a mixture of argon, CO$_{2}$ and heptane (80:20:1.5) gas at standard atmospheric pressure in a pocket with a stainless-steel window. The detection efficiency of this detector is better than 99\% for ions above $\sim$10 MeV/u and a position resolution (FWHM) of 1.9~mm has been reached \cite{Klepper2003}. However, a fraction of electron capture events were lost at the beam energy of 9~MeV/u due to significant energy losses in the pocket window with a thickness of about 25~$\mu$m and the gas with a thickness of around 24~mm before the MWPC.

At about 10 MeV/u, there are only four open nuclear reaction channels: $^{96}$Ru($p, p$)$^{96}$Ru elastic scattering, $^{96}$Ru($p, \gamma$)$^{97}$Rh, $^{96}$Ru($p, \alpha$)$^{93}$Tc and $^{96}$Ru($p, n$)$^{96}$Rh. The above reaction products with different mass-over-charge ratios ($m/q$) were separated by the magnetic field ($B$) of the dipole magnet behind the target. For instance, the dispersion of the magnets displaced $^{97}$Rh$^{45+}$ from unreacted $^{96}$Ru$^{44+}$ by about 142~mm at the position of our detector, see dash-dotted line and solid line in Fig.~\ref{fig:exp_setup_ESR}. As will be shown below, the ($p, \gamma$) reaction products can be discriminated from other nuclear reaction products due to their relatively small momentum spread. The $m/q$ is always larger for atomic EC reaction products since nuclear reaction products are bare. Therefore, orbits of the former are always on the outer side of the ESR, see dashed line in Fig.~\ref{fig:exp_setup_ESR}, and bare nuclear reaction products on the inner side are not contaminated with atomic ones.

The separated nuclear reaction products were detected by the position sensitive double-sided silicon strip detectors (DSSSDs) behind the quadrupole triplet magnets, as shown in Fig.~\ref{fig:exp_setup_ESR}. The DSSSDs consist of two silicon detectors with a $\sim$21.5~mm inactive gap between them, see Fig.~\ref{fig:exp_setup_ESR}. Each silicon detector with an active area of 50$\times$50~mm$^{2}$ has 16 strips in both $x$- and $y$-directions. The strip pitch is 3.1~mm and the strip length is 49.5~mm. The DSSSDs were placed in a pocket separated from the vacuum of the ESR by a 122$\times$44~mm$^{2}$ stainless-steel window with a thickness of about 25~$\mu$m. The thickness of this window limited the reaction energy to above 9~MeV/u in this experiment, since heavy ions with less energy would already be stopped in the window. Recently, an improved detector omitting the window has been mounted \cite{Glorius2014}, which will allow us to measure nuclear reactions around 5~MeV/u in future experiments.

\section{DATA ANALYSIS AND RESULTS}
\subsection{Experimental spectra}
\begin{figure}
\centering
\includegraphics[width=9.2cm]{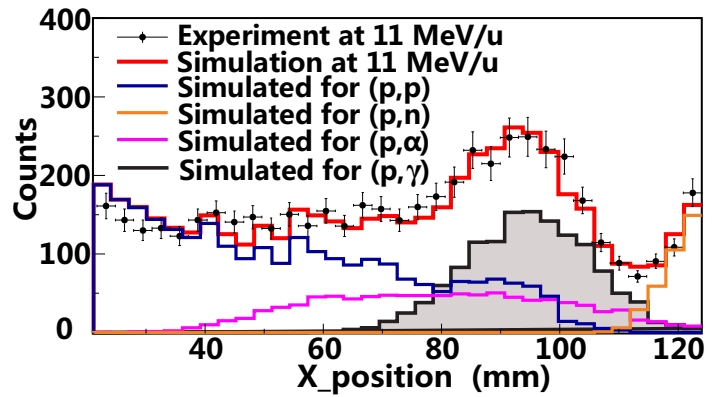}
\caption{(Color online) The experimental $x$ position distribution of nuclear reaction products (black points) registered by the DSSSDs at 11~MeV/u is compared with the simulated $x$ position distribution (red line). Events from different reaction channels, i.e., ($p, p$), ($p, \alpha$), ($p, \gamma$), and ($p, n$) reactions, can be disentangled based on the Geant4 simulation. The ($p, \gamma$) reaction products have a narrower distribution than other nuclear reaction products due to their smaller momentum spread.}
\label{fig:exp_sim_11MeV}
\end{figure}

For each beam energy setting, the DSSSDs were moved to two different positions along the $x$ direction, e.g., 0~mm and 25~mm, to combine two measured spectra into a common $x$ position spectrum without any gap, see black points at 11~MeV/u in Fig.~\ref{fig:exp_sim_11MeV}. To identify the ($p, \gamma$) reaction products unambiguously and to study their transmission efficiency, a Geant4 \cite{Agostinelli2003} simulation has been performed using a numerical model of the experimental setup shown in Fig.~\ref{fig:exp_setup_ESR}. Fig.~\ref{fig:exp_sim_11MeV} compares the simulated total $x$ position spectrum including all nuclear reaction events (red line) with the experimental data (black points) registered by the DSSSDs at 11~MeV/u. The experimental data, which have an uncertainty of about 10\%, can be well reproduced and the ($p, \gamma$) reaction events can be disentangled clearly from the background produced by ($p, p$), ($p, \alpha$), and ($p, n$) reactions, based on Geant4 simulations. The angular distribution of ($p, p$) scattering products in the center-of-mass (CM) system, which serves as an input for the simulation, is from the prediction by TALYS \cite{TALYS_Theta} while other events are assumed to be isotropic. The magnetic fields in the simulation were set to the experimental values. The minimum chi-square method has been utilized to obtain the best simulation spectrum for the experimental data and determine the number of ($p, \gamma$) products $N_{(p,\gamma)}$. The simulation uncertainty can be obtained by simulations varying the sensitive parameters, e.g., magnetic fields, sizes of beam pipes and sizes of chambers, within their uncertainties.

\begin{figure}
\begin{center}
\subfigure {\label{fig:subfig:3a} \includegraphics[width=8.6cm]{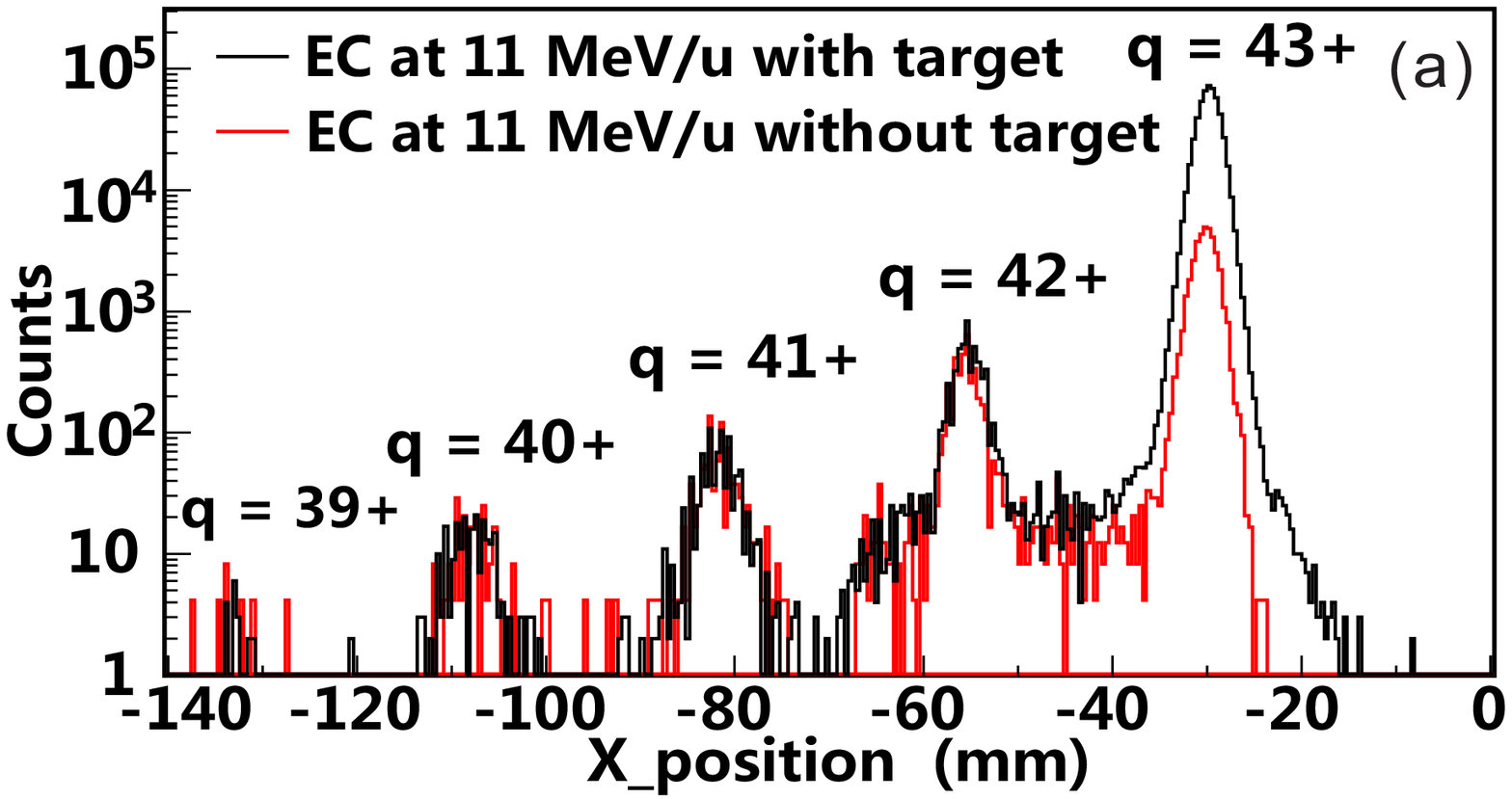}}
\subfigure {\label{fig:subfig:3b} \includegraphics[width=8.65cm]{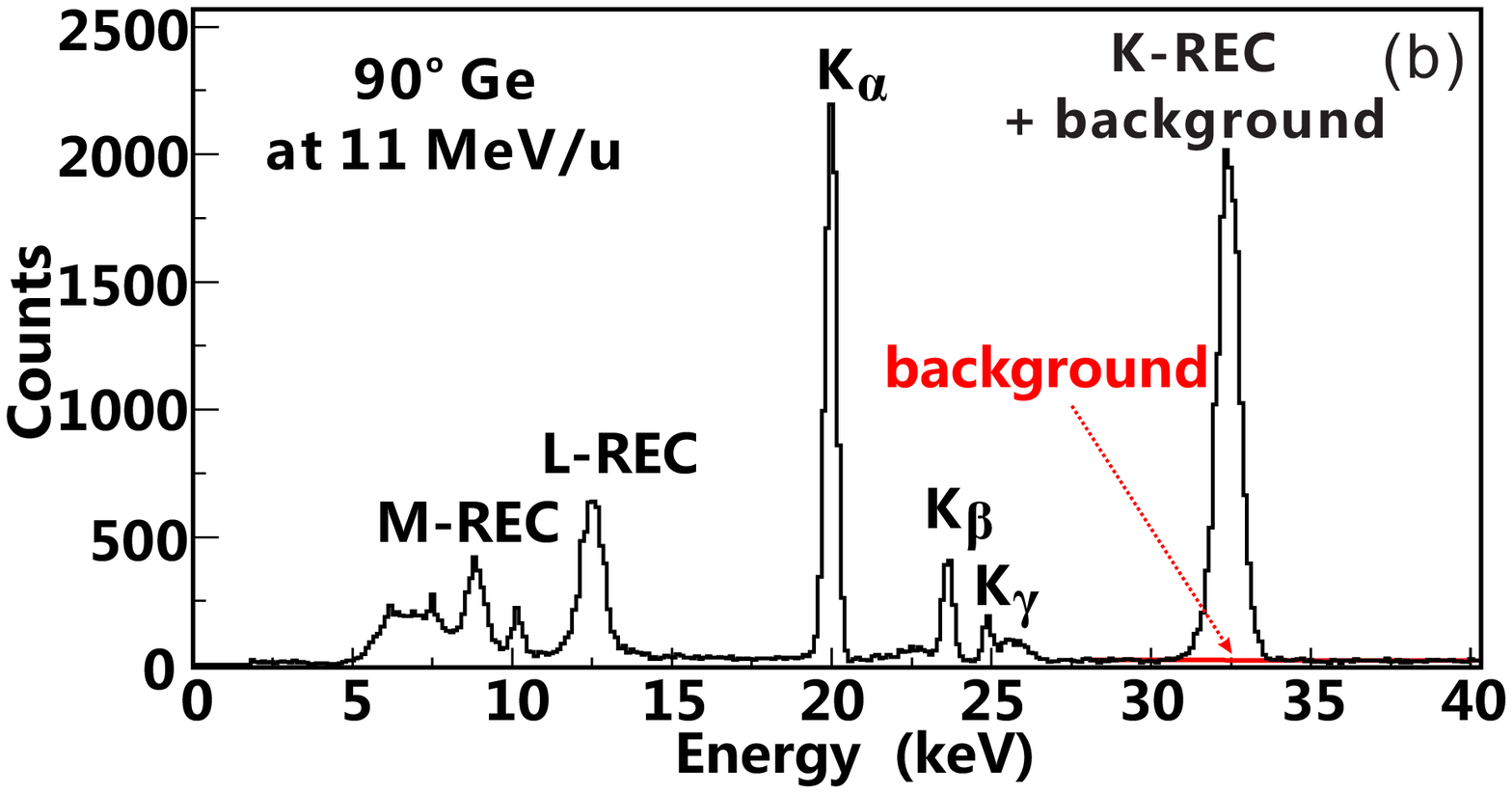}}
\caption{\label{fig:EC+K-REC} (Color online) (a) Position spectra of EC events measured by the MWPC at 11~MeV/u. The black and red lines refer to the spectrum with and without the hydrogen target, respectively. Spectra have been downscaled by a factor of about 300.
(b) X-ray spectrum (black line) at 11~MeV/u registered by the 90$^{\circ}$ Ge detector. The spectrum is not corrected for the detection efficiency. The background is fitted with a linear function.}
\end{center}
\end{figure}

The charge-state spectra measured by the MWPC at 11~MeV/u are presented in Figure~\ref{fig:subfig:3a}. Five different charge states of $^{96}$Ru are produced by the electron capture reactions. The most prominent peak is from $^{96}$Ru$^{43+}$ ions produced by single EC. Other peaks are caused by the capture of more electrons. The black line shows the spectrum measured when the H$_{2}$ target is switched on while the red line indicates the background measured when the H$_{2}$ target is off. Hence, the single EC events caused by the H$_{2}$ target can be determined by subtracting the corresponding background from the interaction of the $^{96}$Ru$^{44+}$ beam with the residual gas in the ring.

The X-ray spectrum measured by the 90$^{\circ}$ Ge detector at 11 MeV/u is given in Figure~\ref{fig:subfig:3b}. The K-, L- and M-REC peaks are caused by the radiative captures into the K-, L- and M-shells of Ru, respectively. The K$_{\alpha}$, K$_{\beta}$ and K$_{\gamma}$ peaks originate from cascades after electron captures into higher shells of Ru. Positions of these peaks are in very good agreement with theoretical predictions of Ref.~\cite{K-REC_E_theory}. According to our experimental data with the H$_{2}$ target switched off, the background spectrum registered by the 90$^{\circ}$ Ge detector is almost linear and there is no peak structure in the K-REC energy region. Therefore, the number of K-REC events induced by the H$_{2}$ target can be extracted from the sum of all events in the K-REC energy region, subtracting a linear background.

\subsection{Cross section determination}
In this experiment, the ($p, \gamma$) cross section can be normalized by two methods using (1) the theoretical single EC cross section, and (2) the calculated K-REC cross section at 90$^{\circ}$. The single EC cross section and the K-REC cross section can be predicted very well by different theoretical models.

For the single EC at about 10~MeV/u, the cross section can be calculated by the Schlachter scaling rule \cite{Schlachter1983}. According to this scaling rule, the cross section of the single EC can be calculated by the relation \cite{Schlachter1983}:
\begin{equation}
\sigma_{\text{EC}} = 1.1\times10^{16}q^{3.9}Z_{2}^{4.2}E^{-4.8}\ \text{barn},
\label{eq:EC_Schlachter}
\end{equation}
where $q$ is the projectile charge state, $Z_{2}$ is 1 for the hydrogen target, and $E$ is the projectile energy in keV/u.
The calculated single EC cross section for our experiment is given in Table~\ref{tab:table1}. Its uncertainty is estimated to be about 20\%. The single EC cross section above 10~MeV/u can also be normalized to the theoretical K-REC cross section, see below, by a validated normalization method reported in Ref.~\cite{Eichler2007} when both single EC and K-REC events were recorded by our detectors. Using this normalization method, the single EC cross section has been calculated by Eq.~(2.2) and Eq.~(2.3) in Ref.~\cite{Eichler2007} (see also Eq.~(\ref{eq:normalization-diff-K-REC}) given below with the ($p, \gamma$) replaced by the single EC). Single EC cross sections determined by two different methods are in good agreement within uncertainties, see Table~\ref{tab:table1}. This agreement also indicates the Schlachter scaling rule works well for $^{96}$Ru$^{44+}$ colliding with H$_{2}$ at about 10~MeV/u in our experiment.
\begin{table}
\caption{\label{tab:table1}Theoretical cross sections of the single EC and corrected differential K-REC cross sections at 90$^{\circ}$ for $^{96}$Ru$^{44+}$ + H$_{2}$ collisions between 9~MeV/u and 11~MeV/u. Single EC cross sections are also normalized to corrected K-REC cross sections at 10~MeV/u and 11~MeV/u.}
\begin{ruledtabular}
\begin{tabular}{cccc}
$E$ (MeV/u)&$\sigma_{\text{EC}}$ (barn)&$\mathrm{d} \sigma_{\text{K-REC}}/\mathrm{d} \Omega$ (barn/sr)&$\sigma_{\text{EC}}^{\text{norm}}$ (barn)\\
\hline 
\noalign{\smallskip}
9&2960$\pm$592&67.06$\pm$13.42&\\
10&1780$\pm$356&59.72$\pm$11.95&1403$\pm$351\\
11&1130$\pm$226&53.74$\pm$10.75&1272$\pm$318\\
\end{tabular}
\end{ruledtabular}
\end{table}

With the number of the ($p, \gamma$) reaction events $N_{(p,\gamma)}$ and the number of the single EC events $N_{\text{EC}}$, the ($p, \gamma$) cross section can be normalized to the single EC cross section $\sigma_{\text{EC}}$, which has been calculated by the scaling rule, using
\begin{equation}
\sigma_{(p,\gamma)} = \frac{N_{(p,\gamma)}}{N_{\text{EC}}} \sigma_{\text{EC}}.
\label{eq:normalization-e-capture}
\end{equation}
$N_{(p,\gamma)}$ and $N_{\text{EC}}$ can be derived from Fig.~\ref{fig:exp_sim_11MeV} and Fig.~\ref{fig:subfig:3a}, respectively. However, the ($p, \gamma$) cross section at 9~MeV/u cannot be determined by this method because some electron capture events produced at this energy were stopped in the pocket window and gas in the pocket before reaching the detector.

Since the K-REC cross section is well-known, the number of K-REC events registered by the 90$^{\circ}$ Ge detector is also used to accurately calculate the ($p, \gamma$) cross section by means of the normalization method. The REC is the inverse of the photoelectric effect, which can be described in the framework of a nonrelativistic dipole approximation by Stobbe \cite{Stobbe1930}. Thus, the total K-REC cross section can be obtained by the principle of detailed balance according to \cite{Eichler2007}
\begin{small}
\begin{equation}
\sigma_{\text{K-REC}}=9165\times\left(\frac{\nu^{3}}{1+\nu^{2}}\right)^{2}\frac{\exp[-4\nu\arctan(1/\nu)]}{1-\exp(-2\pi\nu)}\ \text{barn}.
\label{eq:K-REC_Stobbe}
\end{equation}
\end{small}
$\nu=\alpha Z/\beta$ is the Sommerfeld parameter, $\alpha=1/137.036$ is the fine-structure constant, $\beta=v/c$ is the projectile velocity, and $Z$ is the projectile atomic number. Within the dipole approximation, the differential K-REC cross section at 90$^{\circ}$ can be calculated by the formula \cite{Stohlker1992,Eichler2007}
\begin{equation}
\frac{\mathrm{d} \sigma_{\text{K-REC}}}{\mathrm{d} \Omega} (\theta=90^{\circ})= \frac{3}{8\pi} \sigma_{\text{K-REC}},
\label{eq:Diff-K-REC_Stobbe}
\end{equation}
where $\theta$ is the X-ray observation angle with respect to the beam direction. Then one can calculate the differential K-REC cross section at 90$^{\circ}$ by Eq.~(\ref{eq:K-REC_Stobbe}) and Eq.~(\ref{eq:Diff-K-REC_Stobbe}). However, it has been found that the theoretical value predicted by this formula overestimates the experimental K-REC cross section by a factor of about 30\% \cite{Stohlker1992,Eichler2007}, which appears to be a common feature for K-REC \cite{Eichler2007}. Therefore, an empirical factor of 0.7 for the theoretical prediction was recommended by St{\"o}hlker \emph{et~al.} \cite{Stohlker1992,Eichler2007}. Theoretical cross sections adjusted by this factor agree very well with measured K-REC cross sections, see, e.g., Ref.~\cite{Stohlker1992}. Thus, this factor has also been applied in our calculations of K-REC cross sections. The differential K-REC cross section at 90$^{\circ}$ corrected by this factor is listed in Table~\ref{tab:table1}.

With $N_{(p, \gamma)}$ and the number of K-REC events registered by the 90$^{\circ}$ Ge detector $N_{\text{K-REC}}$, extracted from Fig.~\ref{fig:subfig:3b}, the ($p, \gamma$) cross section can be normalized to the differential K-REC cross section at 90$^{\circ}$ $\mathrm{d} \sigma_{\text{K-REC}}/\mathrm{d} \Omega$ by the following expression
\begin{equation}
\sigma_{(p,\gamma)} = \frac{N_{(p,\gamma)}}{N_{\text{K-REC}}}\varepsilon_{\text{K-REC}} \frac{\mathrm{d} \sigma_{\text{K-REC}}}{\mathrm{d} \Omega}  \Delta\Omega.
\label{eq:normalization-diff-K-REC}
\end{equation}
$\varepsilon_{\text{K-REC}}$ is the intrinsic efficiency and $\Delta \Omega$ is the solid angle spanned by the Ge detector. The K-REC cross section listed in Table~\ref{tab:table1} is adopted to calculate the ($p, \gamma$) cross section using Eq.~(\ref{eq:normalization-diff-K-REC}). Similar methods have been applied in atomic K-REC experiments at the ESR and proved to be justified \cite{Eichler2007}.

\begin{figure}
\centering
\includegraphics[width=8.6cm]{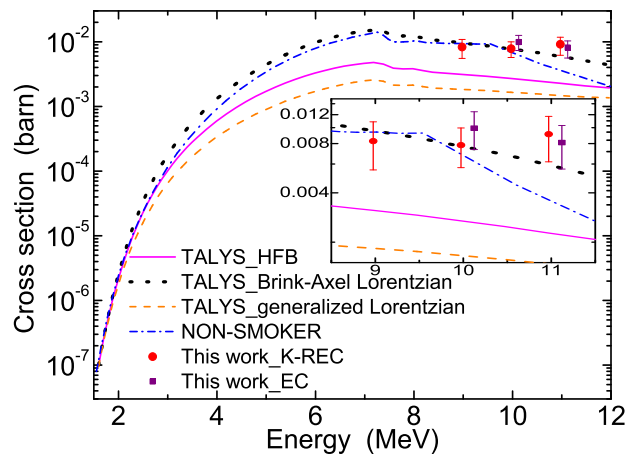}
\caption{(Color online) A comparison of the measured ($p, \gamma$) cross sections with the predictions by the standard NON-SMOKER code \cite{Rauscher2000} (dash-dotted line) as well as the TALYS-1.4 code \cite{Koning2005} using different $\gamma$-ray strength functions, i.e., the HFB (solid line) \cite{Goriely2004}, Brink-Axel Lorentzian (dotted line) \cite{Brink1957,Axel1962} and Kopecky-Uhl generalized Lorentzian (dashed line) \cite{Kopecky1990}. Red points and purple squares represent the measured ($p, \gamma$) cross sections normalized to the K-REC cross sections and the single EC cross sections, respectively. The latter (purple squares) have been slightly offset in energy for clarity. The inset shows a zoom around our measured data. Note that all energies are in the CM system.}
\label{fig:Exp_Gamma_Strength}
\end{figure}
The ($p, \gamma$) cross sections normalized to K-REC cross sections (red points) are in good agreement with those normalized to single EC cross sections (purple squares), as displayed in Fig.~\ref{fig:Exp_Gamma_Strength}.
This agreement gives confidence in the present normalization methods. The uncertainty of the ($p, \gamma$) cross section is dominated by the uncertainty of the calculated K-REC or EC cross section (20\%) and the uncertainty of $N_{(p,\gamma)}$ ($\sim$20\%) while the uncertainty of other factors is always less than 10\%. In the following, ($p, \gamma$) cross sections normalized to K-REC cross sections will be compared with HF calculations since K-REC cross sections have been checked by many atomic experiments and proved to agree well with the experimental values within a small uncertainty of about 20\% \cite{Stohlker1992,Eichler2007}. Another reason is that K-REC events have less background than EC events, see Figs.~\ref{fig:subfig:3a} and \ref{fig:subfig:3b}. The $^{96}$Ru($p, \gamma$)$^{97}$Rh cross section normalized to the K-REC and the associated $S$ factor are summarized in Table~\ref{tab:table2}.
\begin{table}
\caption{\label{tab:table2}Measured cross sections and astrophysical $S$ factors for $^{96}$Ru($p, \gamma$)$^{97}$Rh.}
\begin{ruledtabular}
\begin{tabular}{ccc}
$E_{\text{CM}}$ (MeV)&Cross section (mbarn)&$S$ factor (10$^{8}$ keV barn)\\
\hline 
\noalign{\smallskip}
8.976&8.28$_{-2.76}^{+2.58}$&1.49$_{-0.5}^{+0.47}$\\
9.973&7.83$_{-2.13}^{+2.13}$&0.74$_{-0.2}^{+0.2}$\\
10.971&9.13$_{-2.94}^{+2.59}$&0.5$_{-0.16}^{+0.14}$\\
\end{tabular}
\end{ruledtabular}
\end{table}

\section{MODEL CALCULATIONS}
\subsection{Cross sections}
The $^{96}$Ru($p, \gamma$)$^{97}$Rh cross section is mainly sensitive to the $\gamma$-ray strength function in our experimental energy region ($\sim$10~MeV in the CM system) while it depends sensitively on both the $\gamma$-ray strength function and the proton potential in the Gamow window between roughly 1.3~MeV and 4.3~MeV, see e.g., Fig.~16 in Ref.~\cite{Rauscher2013}. Therefore, our experimental results can be used directly to remove the uncertainty from the $\gamma$-ray strength function and improve the agreement between the theoretical predictions and the experimental data over a large energy region. The comparison of our experimental results by two normalization methods with predictions by NON-SMOKER \cite{Rauscher2000} using default parameters as well as predictions by the TALYS-1.4 code \cite{Koning2005} utilizing different $\gamma$-ray strength functions is presented in Fig.~\ref{fig:Exp_Gamma_Strength}.

The default proton potential from the parameterizations of Koning and Delaroche (KD) \cite{Koning2003} is adopted in TALYS using different $E$1 $\gamma$-ray strength functions, i.e., the Kopecky-Uhl generalized Lorentzian (default strength) \cite{Kopecky1990}, Brink-Axel Lorentzian (BAL) \cite{Brink1957,Axel1962} and microscopic Hartree-Fock-Bogoliubov (HFB) \cite{Goriely2004}.
As shown in Fig.~\ref{fig:Exp_Gamma_Strength}, the ($p, \gamma$) cross section is sensitive to the $\gamma$-ray strength function from about 2~MeV to very high energies and differences between cross sections predicted by TALYS using different $\gamma$-ray strength functions are particularly large at energies above 5~MeV. Therefore, our experimental results between 9~MeV and 11~MeV provide an excellent constraint for the $\gamma$-ray strength function, which can also have a significant impact on the ($p, \gamma$) cross sections in the Gamow window.

The microscopic optical potential of Jeukenne \emph{et~al.} \cite{JLM1977} is applied in the standard NON-SMOKER code. The $\gamma$-ray strength function and nuclear level density models in NON-SMOKER are from Ref.~\cite{Cowan1991} and Ref.~\cite{Rauscher1997}, respectively. The ($p, \gamma$) cross section predicted by NON-SMOKER using the above input has a typical uncertainty of a factor of 2 \cite{Dillmann2011}. Predictions by this code agree well with our experimental results at 9~MeV and 10~MeV. However, NON-SMOKER shows a very drastic tendency of decreasing from 9.5~MeV to higher energies and thus underestimates our data at 11~MeV by a factor of about 2. The possible reason for this discrepancy is that the $\gamma$-ray width is artificially downscaled at high energies to simulate the effect of the pre-equilibrium and in-cascade particle emission in this code \cite{Priv_Rauscher}.

For our experimental results, the best agreement is achieved by TALYS using the BAL $\gamma$-ray strength function. Other models implemented in TALYS severely underestimate our results. A slight deviation between predictions by TALYS using the BAL strength function and our experimental results at 11~MeV may stem from the uncertainty of the proton potential, see below. Considering the good agreement, the BAL strength function will also be applied to predict the ($p, \gamma$) cross sections at lower energies. Note that the uncertainty from the neutron potential has been checked by varying the default KD neutron potential by a factor of $\sim$100 since the ($p, \gamma$) cross section shows a small sensitivity to it. It is found that the variation of the ($p, \gamma$) cross section is very small and thus the uncertainty from the neutron potential does not change the above conclusions. In Fig.~\ref{fig:Exp_Gamma_Strength}, the nuclear level density adopted in TALYS calculations is the widely used back-shifted Fermi gas (BSFG) model \cite{Dilg1973,Koning2008}.

\begin{figure}
\centering
\includegraphics[width=8.6cm]{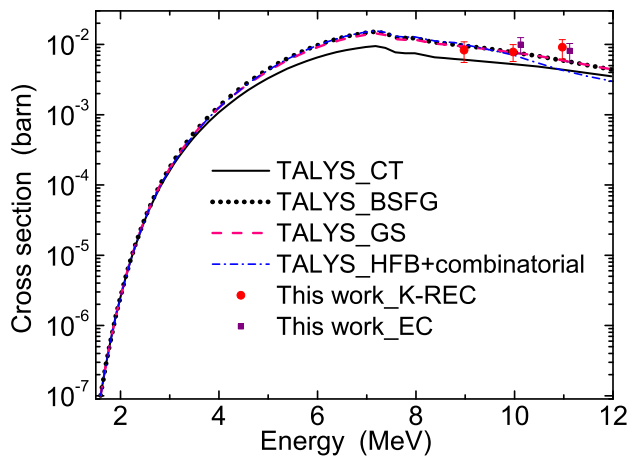}
\caption{(Color online) A comparison of the measured ($p, \gamma$) cross sections with the predictions by the TALYS-1.4 code \cite{Koning2005} using different level density models, i.e., the Constant Temperature (CT) \cite{Gilbert1965}, BSFG \cite{Dilg1973,Koning2008}, Generalized Superfluid (GS) \cite{Ignatyuk1979} and Hartree-Fock-Bogolyubov (HFB) plus combinatorial \cite{Goriely2008} models. The BAL $\gamma$-ray strength function is applied in the TALYS calculations. The purple squares have been slightly offset in energy.}
\label{fig:Exp_NLD}
\end{figure}
The ($p, \gamma$) cross section also shows a relatively small sensitivity to the nuclear level density model. Thus, different level density models, i.e., the Constant Temperature (CT) \cite{Gilbert1965}, BSFG \cite{Dilg1973,Koning2008}, Generalized Superfluid (GS) \cite{Ignatyuk1979} and Hartree-Fock-Bogolyubov (HFB) plus combinatorial \cite{Goriely2008} models, have been tested in TALYS by using the same BAL strength function, as shown in Fig.~\ref{fig:Exp_NLD}. Both the BSFG model and the GS model agree rather well with our experimental data. The relative difference between these two models is always less than 5\%. Therefore, Fig.~\ref{fig:Exp_NLD} also supports our conclusion that TALYS using the BAL $\gamma$-ray strength function as well as the BSFG (or GS) level density model gives the best agreement with our experimental data between 9~MeV and 11~MeV.

After the $\gamma$-ray strength function as well as the level density have been strongly constrained by our experimental data for $^{96}$Ru($p, \gamma$)$^{97}$Rh in the previously unexplored energy region, some additional data for this reaction between 1.6~MeV and 3.4~MeV measured by J. Bork \emph{et~al.} via the activation method \cite{Bork1998} can help to improve the prediction of the $^{96}$Ru($p, \gamma$)$^{97}$Rh cross section further into the Gamow window, where it is sensitive to both the $\gamma$-ray strength function and the proton potential \cite{Rauscher2013}. Recent experimental studies for other proton induced reactions indicate that the proton potential should be modified to explain the measured cross sections \cite{Kiss2007,Kiss2008,Sauerwein2012}. The reason is that the parameters used in the proton potential are usually derived from reactions at energies far above the astrophysical energy region \cite{Rauscher2013}. In the following, the modification of the proton potential will be tested and the uncertainty from the proton potential will be constrained by comparing with experimental results for $^{96}$Ru($p, \gamma$)$^{97}$Rh.

\begin{figure}
\centering
\includegraphics[width=9cm]{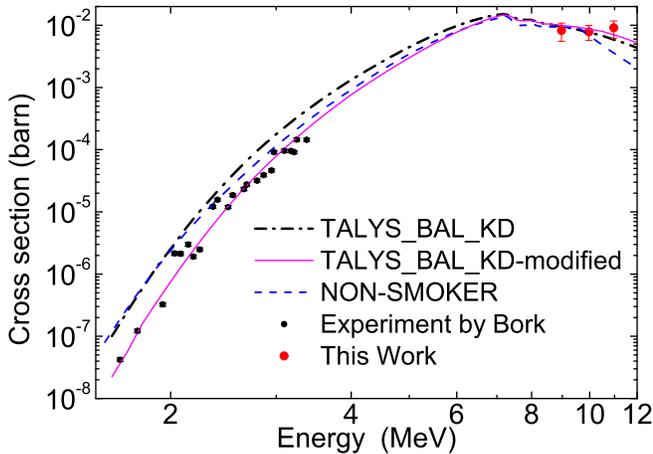}
\caption{(Color online) Our experimental results (red points) and the experimental results of J. Bork \emph{et~al.} (black points) \cite{Bork1998} in different energy regions are compared with predictions by NON-SMOKER \cite{Rauscher2000} (blue dashed line) and TALYS using the KD proton potential (black dash-dotted line) as well as TALYS using the modified KD proton potential (magenta solid line). The staggering structure in the low energy experimental data \cite{Bork1998} is due to the low level density. Note that both the cross section and the energy are plotted on a logarithmic scale.}
\label{fig:Exp_Model_OMP}
\end{figure}
In Fig.~\ref{fig:Exp_Model_OMP}, $^{96}$Ru($p, \gamma$)$^{97}$Rh cross sections from two experiments in different energy regions are compared with predictions by NON-SMOKER \cite{Rauscher2000} and TALYS \cite{Koning2005} using the standard KD proton potential as well as TALYS using the modified KD proton potential.
In the modified KD proton potential, default values of $r_{V}$ and $d_{1}$ have been scaled by factors of 0.7 and 0.8, respectively, to obtain a good agreement with the data measured by Bork \emph{et~al.} \cite{Bork1998} and also our new data. The BAL $\gamma$-ray strength function and the BSFG level density are utilized in TALYS because they can provide the best prediction for our results between 9~MeV and 11~MeV, see Figs.~\ref{fig:Exp_Gamma_Strength} and \ref{fig:Exp_NLD}. Both NON-SMOKER and TALYS using the standard KD proton potential significantly overestimate cross sections at low energies, e.g., by a factor of around 5 at energies below 2~MeV. TALYS using the BAL $\gamma$-ray strength function, the BSFG level density and modified KD proton potential can excellently reproduce both experimental results in different energy regions. It is remarkable to see that TALYS using constrained parameters can reproduce cross sections varying from about 3$\times$10$^{-8}$~barn to $\sim$1$\times$10$^{-2}$~barn in a large energy range from about 1.5~MeV to 11~MeV.

\subsection{Reaction rates}
\begin{figure}
\centering
\includegraphics[width=9cm]{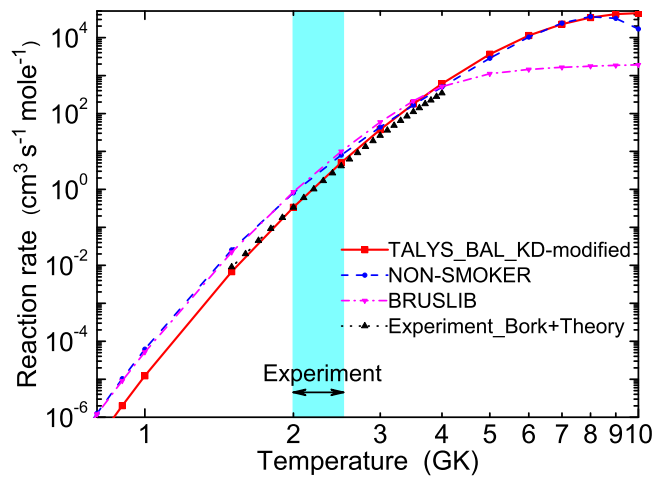}
\caption{(Color online) Stellar reaction rates over a large temperature region calculated by the TALYS code using the modified KD proton potential (red solid line). They are compared to rates from the experimental data of Bork \emph{et~al.} \cite{Bork1998} (black triangles in the filled region) as well as the theoretical extrapolation (black triangles out of the filled region), NON-SMOKER calculations \cite{Rauscher2000} (blue dashed line) and BRUSLIB \cite{BRUSLIB} (magenta dash-dotted line). Both the rate and the temperature are illustrated on a logarithmic scale.}
\label{fig:Exp_Model_Reaction_Rates}
\end{figure}
As an input for $p$-process network calculations, the stellar rate $N_{A}\langle\sigma^{*}\upsilon\rangle_{(p, \gamma)}$ over a large temperature range, especially in the astrophysically relevant temperature range between 1.5~GK and 3.5~GK, is required \cite{Arnould2003,Rauscher2013}. However, the $^{96}$Ru($p, \gamma$)$^{97}$Rh cross section between 1.6~MeV and 3.4~MeV measured by J. Bork \emph{et~al.} is only sufficient to compute the rate between about 2~GK and 2.5~GK \cite{Bork1998}, see black triangles in the filled region in Fig.~\ref{fig:Exp_Model_Reaction_Rates}. Experimental data must be extended to a wider energy region by HF calculations. The TALYS cross section constrained by two experimental data sets, as presented in the previous section, represents a solid basis for the reaction rate calculation covering the important temperature range.

The stellar reaction rates calculated by the TALYS code using constrained nuclear physics input are compared to the experimental rates around 2~GK from Ref.~\cite{Bork1998} as well as their theoretical extrapolation, NON-SMOKER calculations \cite{Rauscher2000} and BRUSLIB \cite{BRUSLIB}, as presented in Fig.~\ref{fig:Exp_Model_Reaction_Rates}.
\begin{table}
\caption{\label{tab:table3}Recommended stellar rates at different temperatures for $^{96}$Ru($p, \gamma$)$^{97}$Rh. The rate has an uncertainty of about 30\% according to the comparison with measured data.}
\begin{ruledtabular}
\begin{tabular}{ccc}
Temperature (GK)&Stellar rate (cm$^{3}$s$^{-1}$mole$^{-1}$)\\
\hline 
\noalign{\smallskip}
0.5&(1.8$\pm$0.5)$\times$10$^{-11}$\\
0.6&(9.1$\pm$2.7)$\times$10$^{-10}$\\
0.7&(2.0$\pm$0.6)$\times$10$^{-8}$\\
0.8&(2.5$\pm$0.8)$\times$10$^{-7}$\\
0.9&(2.1$\pm$0.6)$\times$10$^{-6}$\\
1&(1.3$\pm$0.4)$\times$10$^{-5}$\\
1.5&(7.0$\pm$2.1)$\times$10$^{-3}$\\
2&(3.4$\pm$1)$\times$10$^{-1}$\\
2.5&5.0$\pm$1.5\\
3&(3.7$\pm$1.1)$\times$10$^{1}$\\
3.5&(1.8$\pm$0.5)$\times$10$^{2}$\\
4&(6.1$\pm$1.8)$\times$10$^{2}$\\
5&(3.7$\pm$1.1)$\times$10$^{3}$\\
6&(1.1$\pm$0.3)$\times$10$^{4}$\\
7&(2.3$\pm$0.7)$\times$10$^{4}$\\
8&(3.4$\pm$1)$\times$10$^{4}$\\
9&(4.1$\pm$1.2)$\times$10$^{4}$\\
10&(4.4$\pm$1.3)$\times$10$^{4}$\\
\end{tabular}
\end{ruledtabular}
\end{table}
\begin{table*}
\caption{\label{tab:table4}Recommended REACLIB parameters for the reactivity of $^{96}$Ru($p, \gamma$)$^{97}$Rh and its reverse reaction.}
\begin{ruledtabular}
\begin{tabular}{cccccccc}
&$a_{0}$&$a_{1}$&$a_{2}$&$a_{3}$&$a_{4}$&$a_{5}$&$a_{6}$\\
\hline 
\noalign{\smallskip}
($p, \gamma$)&0.90232&-10.08917&-5.75549&2.62112&1.01194&-0.29641&4.64569\\
($\gamma, p$)&22.29232&-54.302315&-5.75549&2.62112&1.01194&-0.29641&6.14569\\
\end{tabular}
\end{ruledtabular}
\end{table*}
The experimental rates have been theoretically extrapolated by normalizing the $^{96}$Ru($p, \gamma$)$^{97}$Rh cross sections calculated by NON-SMOKER to the experimental data of J. Bork \emph{et~al.}, see Ref.~\cite{Bork1998}. In the normalization, a factor of about 0.5 has been applied for NON-SMOKER calculations. The TALYS rate using parameters constrained in the previous section can excellently reproduce the experimental rate for $^{96}$Ru($p, \gamma$)$^{97}$Rh between 2~GK and 2.5~GK, as displayed in Fig.~\ref{fig:Exp_Model_Reaction_Rates}. Hence, the TALYS rate constrained by two experimental data sets at different energy regions is recommended for $^{96}$Ru($p, \gamma$)$^{97}$Rh, as listed in Table~\ref{tab:table3}.

On the contrary, both NON-SMOKER and BRUSLIB overestimate the rate at temperatures below 3~GK, particularly at low temperatures. A good agreement is reached between predictions by NON-SMOKER and TALYS between 3.5~GK and 9~GK. However, NON-SMOKER underestimates the rate above 9~GK, which is caused by the underestimation of the cross section above about 10~MeV. Above 3~GK, the rate extrapolated by NON-SMOKER is lower than the TALYS rate since the former are determined by normalizing the NON-SMOKER rate to the experimental rate around 2~GK. In addition, BRUSLIB significantly underestimates the rate above 4~GK. For instance, BRUSLIB underestimates the rate by a factor of 20 above 8~GK, which again indicates that our experimental data are important to constrain the theoretical rate. It should be stressed that the BRUSLIB rate is calculated by the TALYS code, where the input parameters are not constrained by experiments, see Ref.~\cite{BRUSLIB} for details.

For network calculations, the recommended rate has been parameterized in the REACLIB format \cite{Rauscher2000} using the formula
\begin{eqnarray}
N_{A}\langle\sigma\upsilon\rangle = \exp[a_{0}+a_{1}T_{9}^{-1}+a_{2}T_{9}^{-1/3}+a_{3}T_{9}^{1/3}\nonumber\\
+a_{4}T_{9}+a_{5}T_{9}^{5/3}+a_{6}\ln\left(T_{9}\right)],
\label{eq:rate_fit}
\end{eqnarray}
where $T_{9}$ is the temperature in GK. Recommended REACLIB parameters are listed in Table~\ref{tab:table4}. The fit using these parameters agrees very well with the recommended rate within 10\% between 1~GK and 8~GK.

\section{SUMMARY AND OUTLOOK}
In summary, a novel technique via the collision of stored heavy ions with a hydrogen target has been developed at the ESR, which provides unrivalled opportunities for the direct measurement of ($p, \gamma$) reactions around the energy range of astrophysical interest, particularly for previously unreachable radioactive ions.
This method has been successfully demonstrated for the first time by measuring the $^{96}$Ru($p, \gamma$)$^{97}$Rh cross sections between 9~MeV and 11~MeV. The present experimental results allowed us to pin down the $\gamma$-ray strength function, which is a critical parameter in the HF model, as well as the nuclear level density model. After this, another important parameter, the proton potential, has also been constrained by combining our results with some additional data at lower energies. TALYS, constrained by two experiments in different energy regions, can excellently predict the stellar rates for $^{96}$Ru($p, \gamma$)$^{97}$Rh over a large temperature range for $p$-process network calculations.

Further measurements of ($p, \gamma$) reactions at lower energies around the Gamow window via this method using our improved detector are planned in future experiments at heavy ion storage rings. Besides, ($\alpha, \gamma$) reactions will also be measured by this method when a helium target is utilized.

\begin{acknowledgments}
We thank Dr. T. Rauscher for his help on the NON-SMOKER code. This research was partially supported by HIC for FAIR, NAVI and the HYG PIANO.
\end{acknowledgments}

\providecommand{\noopsort}[1]{}\providecommand{\singleletter}[1]{#1}%

\end{document}